\documentclass[aip, amsmath, amssymb, reprint]{revtex4-1}
\usepackage{bbold}
\usepackage[dvipsnames]{xcolor}
\usepackage{dcolumn}
\usepackage{bm}
\usepackage[utf8]{inputenc}
\usepackage[T1]{fontenc}
\usepackage{mathptmx}
\usepackage{etoolbox}
\usepackage{caption}
\usepackage{subcaption}
\usepackage{graphicx}
\usepackage{float}

\makeatletter
\def\@email#1#2{
 \endgroup
 \patchcmd{\titleblock@produce}
  {\frontmatter@RRAPformat}
  {\frontmatter@RRAPformat{\produce@RRAP{*#1\href{mailto:#2}{#2}}}\frontmatter@RRAPformat}
  {}{}
}
\makeatother

\begin{document}

\preprint{AIP/123-QED}

\title{Discretizing the Bistability of Mode-Locked Electron Spin Precession: An Overhauser Field Hysteresis Manifestation}
\author{Estefanio Kesto}
 \homepage{Electronic mail: ekesto@umich.edu}
 \author{Michael J. Dominguez}
\author{Vanessa Sih}
\affiliation{ 
Department of Physics, University of Michigan, Ann Arbor, Michigan 48109, USA
}

\date{\today}

\begin{abstract}
Electron-nuclear spin interactions by pulsed optical pumping have been found to polarize the nuclear spin system, leading to the nuclei building up an intrinsic magnetic field known as the Overhauser field. Studies have indicated an Overhauser field hysteresis effect dependent on the sweep direction of an externally applied magnetic field in negatively detuned periodically pumped Si-doped GaAs. Although predictions of bistable mode-locked electron spin precession frequency modes have been made for systems exhibiting this hysteresis, there have been no reports on the experimental observation of said bistable spin precession modes. This letter details the evolution of bistable Overhauser field solutions leading to a hysteretic effect in negatively detuned optical excitation of Si-doped GaAs by magneto-optic pump-probe spectroscopy and investigates the resulting consequence of this hysteresis acting on the electron spin system. One manifestation of the Overhauser field hysteresis acting on the electron spin system leads to the discretization of bistable mode-locked electron spin precession modes within a given band of externally applied magnetic field. A method for preferentially accessing the two different and stable mode-locked spin precession modes within a given band of externally applied magnetic field is outlined, which may be of interest for communities utilizing electron and nuclear spins for information processing protocols.
\end{abstract}

\maketitle

Early experimental investigations in GaAs quantum wells have shown that detuning the laser source to a nonresonant wavelength results in sufficient optical Stark shift \cite{OSE1}, such that the optical Stark effect (OSE) may be experimentally observed. Subsequent theoretical and experimental analyses of the OSE in bulk GaAs at low temperatures concluded that the OSE is also present in bulk materials \cite{OSE2, OSE3}. The underlying mechanism allowing the OSE to be observed in bulk GaAs at low temperatures was attributed to suppressing thermal ionization of excitons, such that the exciton linewidth of bulk GaAs converges with that of GaAs quantum wells \cite{OSE4, OSE6, OSE5}.

More recent studies on the OSE done in the Voigt geometry found that detuning the laser from a materials resonant energy causes an electron spin rotation, thereby producing electron spin polarization parallel or antiparallel to the externally applied magnetic field (${B}_{ext}$) \cite{RotateSpin1, RotateSpin2}. A result of producing electron spin polarization parallel or antiparallel to ${B}_{ext}$ is the generation of strong dynamic nuclear polarization (DNP) and a non-negligible Overhauser field (${B}_{N}$) that interacts with the electron spin system \cite{PumpProbe1}. Generating strong DNP and building up a non-negligible ${B}_{N}$ by means of negative detuning has also been found to generate electrons with mode-locked precession frequencies in well-defined bands of ${B}_{ext}$ \cite{Discretize}. This effect, known as nuclear-induced frequency focusing (NIFF) \cite{PumpProbe1, NIFF1, NIFF2}, results from the discretization of the effective effective magnetic field experienced by the electrons \cite{Discretize}.

The demonstration of discretizing the effective magnetic field that the electrons experience in Ref. \citenum{Discretize} was investigated in ZnSe using pump-probe spectroscopy. The Overhauser field acting together with ${B}_{ext}$ is what led to the discretized effective magnetic field and, consequently, the locking of the electron spin precession frequency into particular modes in bands of ${B}_{ext}$ \cite{Discretize}. However, the system under investigation in Ref. \citenum{Discretize} did not find the Overhauser field in their periodically pumped ZnSe system having hysteretic features.

Should a system under investigation have the presence of an Overhauser field hysteresis, then the electron spin system may be able to experience two different discretized effective magnetic fields in the same band of $B_{ext}$. The effective magnetic field (${B}_{eff}$) and effective angular precession frequency (${\Omega}_{eff}$) that the spin-polarized electrons experience are defined by

\begin{subequations}\label{eq:a}
        \begin{equation}\label{eq:a1}
            \overrightarrow{B}_{eff}=\overrightarrow{B}_{ext}+\overrightarrow{B}_N
        \end{equation}
        
        \begin{equation}\label{eq:a2}
            {\Omega}_{eff} = g{\mu}_BB_{eff}/{\hbar} = g{\mu}_B(B_{ext} + B_{N})/{\hbar}
        \end{equation}
\end{subequations}

\noindent where $g$ is the electron g-factor, ${\mu}_B$ is the Bohr magneton, and $\hbar$ is reduced Planck's constant. Considering that ${\Omega}_{eff}$ depends on $B_N$ through $B_{eff}$, then the electrons may have different ${\Omega}_{eff}$ in the same band of $B_{ext}$. This condition of attaining two different ${\Omega}_{eff}$ in the same band of $B_{ext}$ requires being able to reach the higher and lower bistable Overhauser field solutions, which may be revealed through investigating hysteresis effects on the Overhauser field.

Investigations on bistable Overhauser field solutions leading to a hysteretic effect have studied its dependence on power and polarization from the excitation source \cite{OverhauserHysteresis2, OverhauserHysteresis3, OverhauserHysteresis4}. Further investigations have studied field-dependent effects on the electron-nuclear spin dynamics, thereby confirming the existence of an Overhauser field hysteresis \cite{PL2,fielddepend, PumpProbe1}, and subsequent theoretical predictions of overlapped (i.e., bistable) spin precession modes mediated through the Overhauser field hysteresis have been made \cite{OverhauserHysteresis6, OverhauserHysteresis7}. While prior studies have theoretically predicted bistable spin precession modes for systems exhibiting a hysteretic Overhauser field, experimental verification has not yet been reported.
 
This letter experimentally and computationally demonstrates that negatively detuned periodic optical pumping of Si-doped GaAs by magneto-optic pump-probe spectroscopy induces an Overhauser field hysteresis that gives rise to bistable mode-locked spin precession modes existing in the same band of ${B}_{ext}$, thereby experimentally proving the prediction made by Ref. \citenum{OverhauserHysteresis6}. More specifically, this letter details: (1) experimentally mapping out the evolution of the Overhauser field latching onto the different theoretical bistable solutions and the resulting discretization of the bistable $B_{eff}$; (2) a hysteretic behavior in the electron spin precession frequency leading to different spin precession modes in the same band of $B_{ext}$; and (3) the discretization and mode-locking of the different spin precession modes. Combined, this work demonstrates a method for preferentially accessing the different mode-locked spin precession frequency modes existing in the same band of $B_{ext}$.

A GaAs epilayer with a thickness of 2 microns and active Si-doped density of $n=3 \times 10^{16}$ cm$^{-3}$ was used for this study. The substrate used to grow the active GaAs is a 1-micron-thick undoped AlGaAs grown on top of an undoped (001)-oriented GaAs wafer. Both the active GaAs and inactive AlGaAs are grown by molecular-beam epitaxy. A helium flow cryostat placed in between the poles of an electromagnet is utilized to maintain a nominal sample temperature of 10 K.

\begin{figure*}
     \hspace*{-1.5cm}\includegraphics[scale = 0.5]{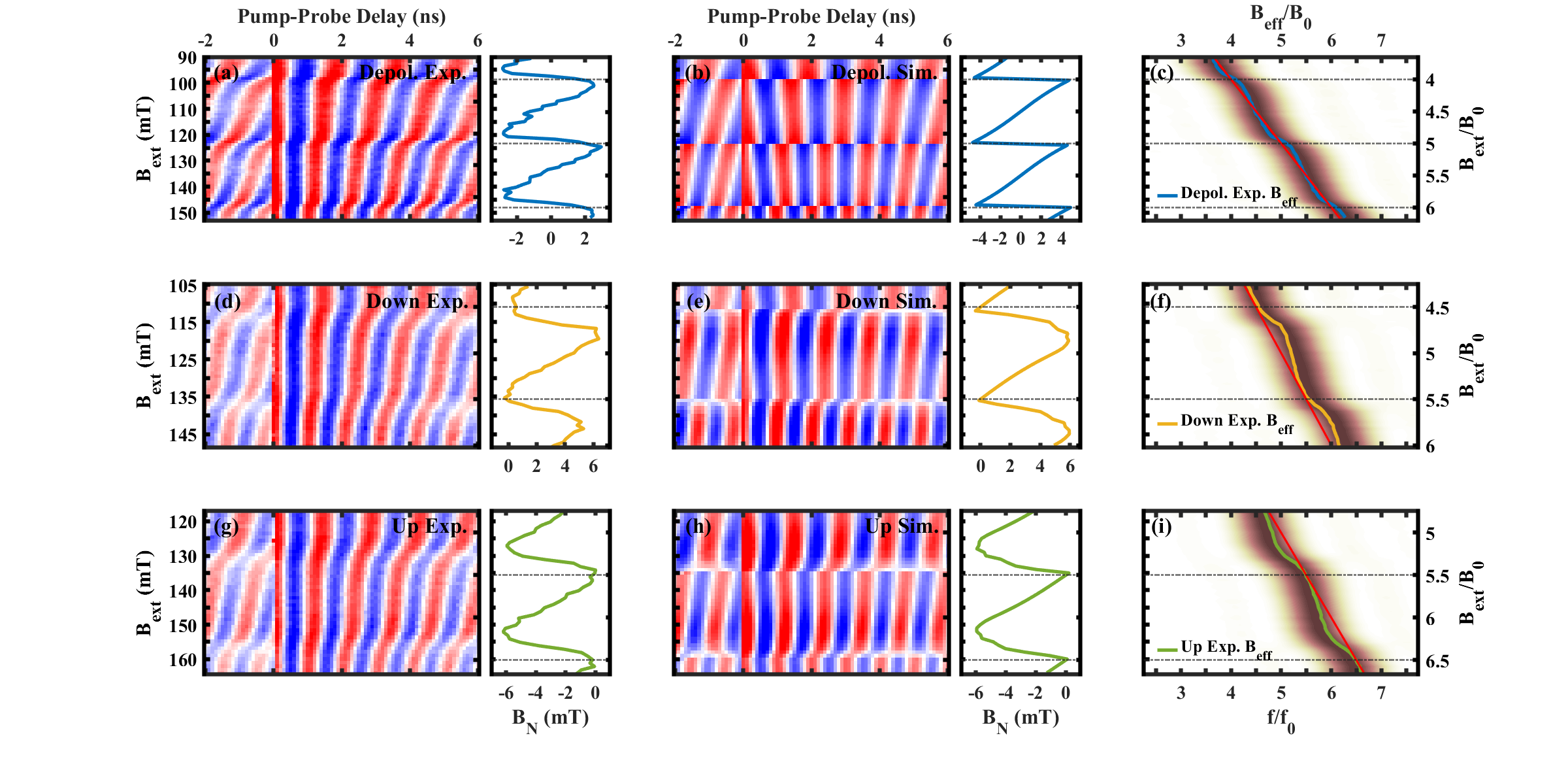} 
    \caption{Discretization of the bistable mode-locked electron spin precession frequencies. Figures (a)--(c), (d)--(f), and (g)--(i) correspond to the method of depolarizing, downsweeping, and upsweeping ${B}_{ext}$, respectively. Figures (a)/(b), (d)/(e), and (g)/(h) present experimental/simulated 2D TRKR and FRKR color plots of $S_z(t)$, with the panel to the right of the color plot corresponding to the respective Overhauser field. Figures (c), (f), and (i) are 2D color plots of the normalized powers of a fast Fourier transform (FFT) applied to the experimental $S_z(t)$ color plots with their corresponding $B_{eff} = B_{ext} + B_N$ overlaid on them. The horizontal dashed lines and red lines depict the locations associated with the switching of precession frequency and the results expected for no Overhauser field, respectively.} 
    {\phantomsubcaption\label{fig:2a}
     \phantomsubcaption\label{fig:2b}
     \phantomsubcaption\label{fig:2c}
     \phantomsubcaption\label{fig:2d}
     \phantomsubcaption\label{fig:2e}
     \phantomsubcaption\label{fig:2f}
     \phantomsubcaption\label{fig:2g}
     \phantomsubcaption\label{fig:2h}
     \phantomsubcaption\label{fig:2i}
     }
    \label{fig:2}
\end{figure*}

Time- and field-resolved Kerr rotation (TRKR and FRKR) pump-probe spectroscopy in the Voigt configuration is utilized to generate and detect the spin dynamics. The TRKR/FRKR measurements polarize the electron spin system with a right circularly-polarized (RCP) pump beam and the spin dynamics are detected using a linearly-polarized (LP) probe beam. The pulses are generated by a mode-locked Ti:S laser configured to output pulses with a 2-ps duration, and the experimental apparatus configures the pump pulse repetition period to be 6.58 ns.

Previous investigations operating at the same temperature and utilizing the same Si-doped GaAs has found the maximum Overhauser field to occur at an operating wavelength of $\sim$819.5 nm \cite{fielddepend, PumpProbe1}. Therefore, the Ti:S source utilized for this investigation is negatively-detuned to $\sim$819.5 nm in order to generate a sufficient Overhauser field for NIFF to occur. The Ti:S source used for this study has an intrinsic finite-spectral bandwidth with a full-width at half maximum (FWHM) of $\sim$0.6 nm. The $\sim$0.6 nm FWHM results in an appreciable spread in the optical Stark rotation \cite{PumpProbe2}, thereby resulting in a non-negligible distribution of Overhauser fields. The finite spectral bandwidth of $\sim$0.6 nm FWHM is implemented into the simulated TRKR/FRKR results to gain a full picture of how the distributions of the optical Stark rotation and Overhauser field effects the spin dynamics. 

The FRKR measurements are done using three separate methods: (1 - Depolarize) sweeping ${B}_{ext}$ from 90 mT to 153 mT in steps of 1 mT, where the electromagnet is set to 300 mT for $\sim$1 s, then set to 0 mT for $\sim$1 s, and then set to -300 mT for $\sim$1 s before setting the corresponding field in the measurements; (2 - Downsweep) sweeping ${B}_{ext}$ from 150 mT to 104 mT in steps of 1 mT; and (3 - Upsweep) sweeping ${B}_{ext}$ from 116 mT to 165 mT in steps of 1 mT. The "depolarize" method gets its name because setting the electromagnet to 300 mT for $\sim$1 s, then to 0 mT for $\sim$1 s, and then to -300 mT for $\sim$1 s before setting the corresponding field completely depolarizes the nuclear spin system and, consequently, resets the Overhauser field (i.e., ${B}_{N} = 0$) at the start of each TRKR measurement \cite{Depolarize1, param2}. After the desired field is set for each respective method, the pump-probe time delay (${\Delta}t$) is swept from -2000 ps to 6000 ps in steps of 100 ps with $\sim$4.5 s of lab time between each step for the TRKR measurements. The data from the TRKR/FRKR measurements are then aggregated into a 2D color plot (see Fig. \ref{fig:2}) giving insight into the spin dynamics by observing changes in the NIFF effects and the corresponding Overhauser fields.

The experimentally extracted Overhauser fields presented in this report are measured by curve fitting the TRKR measurements for positive time delay at their discrete ${B}_{ext}$ to a decaying cosine having the form

\begin{equation}\label{eq:4}
    {\theta}_k \propto S_z(t) \propto A \text{cos}({\Omega}_{eff}{\Delta}t)e^{^{-{\Delta}t}/_{T_2^*}}
\end{equation}

\noindent where $S_z(t)$ is the z-component of electron spin polarization and A is the amplitude of the Kerr rotation angle (${\theta}_k$). The curve fitting process extracts the effective angular precession frequency (Eq. \ref{eq:a2}) at each discrete ${B}_{ext}$, then rearranges it in the form (Eq. \ref{eq:5})

\begin{equation}\label{eq:5}
    \overrightarrow{B}_N = \Omega_{eff}\frac{\hbar}{g\mu_B} - \overrightarrow{B}_{ext}
\end{equation}

\noindent to solve for ${B}_N$.

The simulated spin dynamics presented in this report are modeled by assuming an infinite train of negatively detuned pump pulses separated by a repetition period of 6.58 ns. Due to the DNP process, the buildup of nuclear spin polarization, which the electron spin components depend on, are modeled using a nuclear rate equation having the form

\begin{equation}\label{eq:2a}
    \frac{dI_{av}}{dt} = -\frac{1}{T_{1e}} \left(  I_{av}  - \frac{4}{3}I(I+1) \langle S_x \rangle  \right) -\frac{I_{av}}{T_{1n}}
\end{equation}

\noindent where $I_{av}$ is the fraction of nuclear spin polarization, $I$ is the nuclear spin value for Ga and As isotopes, $T_{1e}$ is the nuclear polarization time, $T_{1n}$ is the phenomenological term incorporating all nuclear relaxation mechanisms, and $\langle S_x \rangle$ is the time-averaged x-component of the electron spin polarization. The computational model used to simulate the spin dynamics presented herein were adopted from Refs. \citenum{PumpProbe1} and \citenum{PumpProbe2}, and the model produced by Ref. \citenum{OverhauserHysteresis6} may also be used to simulate these respective spin dynamics. Thorough details regarding the computational model, all relevant equations, and the corresponding values for the parameters utilized for the simulations presented herein can be found in supplementary note 1.

\begin{figure}
    \includegraphics[width=1\columnwidth]{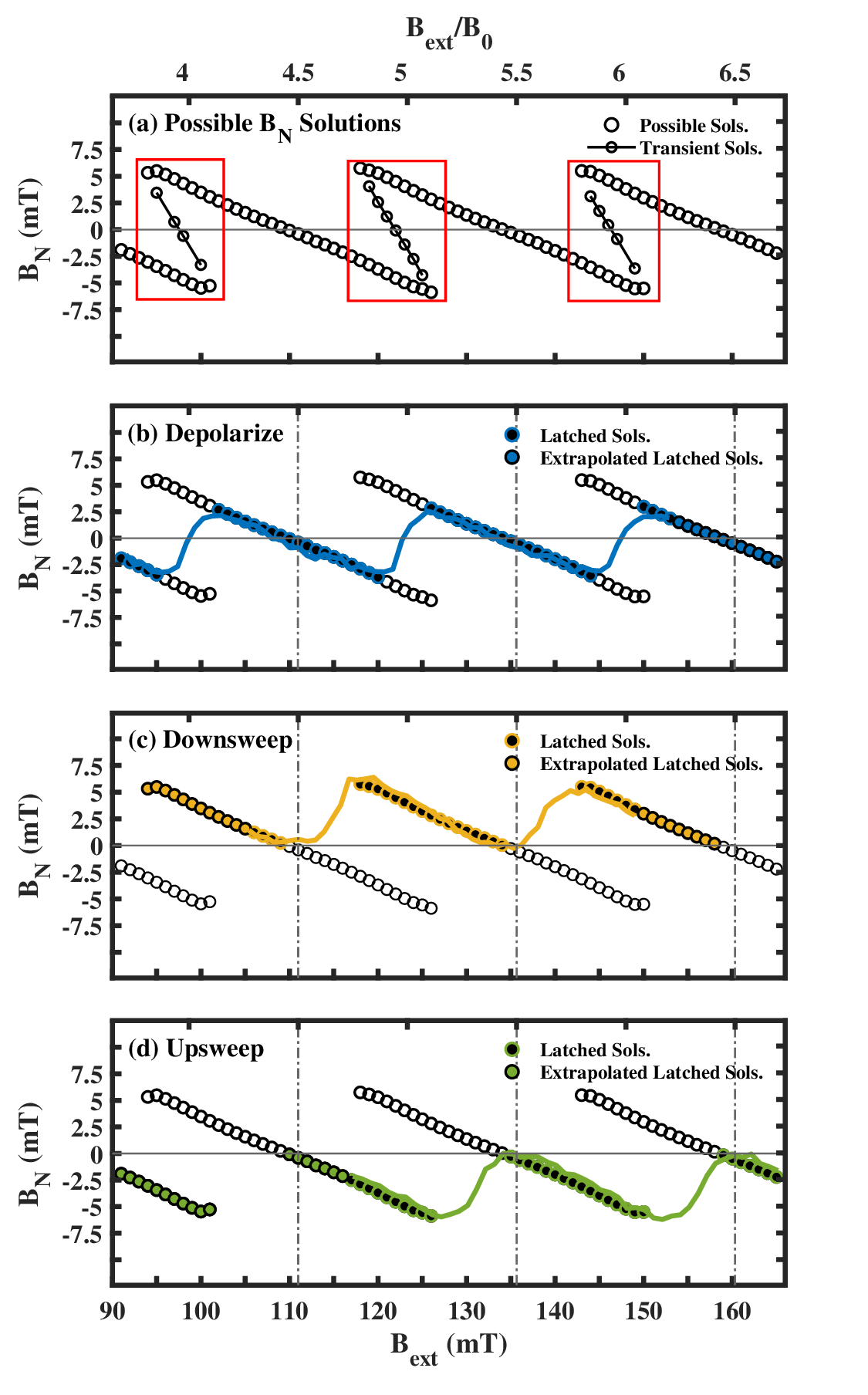}
    \caption{Overhauser field ${B}_N$ plotted as a function of externally applied magnetic field ${B}_{ext}$. (a) Computationally produced Overhauser field solutions, where the bistable solutions are enclosed in red boxes for clarity. (b)/(c)/(d) Depolarized/Downsweep/Upsweep experimentally extracted ${B}_N$ with overlaid stable ${B}_N$ solutions. The black and colored filled circles represent the latched $B_N$ solutions and the extrapolation used in Fig. \ref{fig:4}, respectively.}
    {\phantomsubcaption\label{fig:3a}
    \phantomsubcaption\label{fig:3b}
     \phantomsubcaption\label{fig:3c}
     \phantomsubcaption\label{fig:3d}
     }
    \label{fig:3}
\end{figure}

Figure \ref{fig:2} presents the discretization of bistable electron mode-locked spin precession frequencies as a result of the nuclei building up a non-negligible Overhauser field (see the supplementary note 1 for a simulation of the spin dynamics with no presence of an Overhauser field). The Overhauser field build-up provides some level of discretization of the effective magnetic field and, subsequently, leads to some level of a discretized precession frequency \cite{Discretize}. The discretized magnetic field and precession frequency are given by

\begin{subequations}\label{eq:6}
    \begin{equation}\label{eq:6a}
        B_{eff} \rightarrow B_{k} = k\frac{2{\pi}{\hbar}}{|g|{\mu}_BT_{rep}} = kB_0
    \end{equation}
    \begin{equation}\label{eq:6b}
        f_{k} = \frac{{\Omega}_{eff}}{2\pi} = B_{k}\frac{|g|{\mu}_B}{2\pi\hbar}=k\frac{1}{T_{rep}} = kf_0
    \end{equation}
\end{subequations} 

\noindent where k is an integer, $B_{0} \approx 24.6$ mT, and $f_{0} \approx 0.153$ GHz for this system.

The computed solutions to the Overhauser field are presented in Fig. \ref{fig:3a}, which shows that there may be one or more possible solutions for the Overhauser field at a given $B_{ext}$ (i.e., there exists bistable Overhauser field solutions). The corresponding Overhauser field trajectories for depolarizing, downsweeping, and upsweeping $B_{ext}$ are shown as solid lines in Figs. \ref{fig:3b}, \ref{fig:3c}, and \ref{fig:3d}, respectively. 

The Overhauser field trajectory for the depolarizing method latches onto a stable ${B}_N$ solution that is closest to zero at each $B_{ext}$, which is shown in Fig. \ref{fig:3b}, rather than continuing to follow the trajectory of the stable solutions. This is due to configuring $B_{ext}$ to depolarize the nuclear spin system prior to the start of each TRKR scan and, consequently, ${B}_N = 0$ prior to the start of each TRKR scan. However, the downsweep and upsweep methods do not latch onto the stable ${B}_N$ solution that is closest to zero at each $B_{ext}$.

For downsweeping and upsweeping ${B}_{ext}$, ${B}_N$ can be seen to follow the trajectory of the stable solutions, which are shown in Figs. \ref{fig:3c} and \ref{fig:3d}, respectively. More specifically, the downsweeping (upsweeping) scan start at a $B_{ext}$ of 150 (116) mT, which forces the initial latched solution to be positive (negative). Then, as $B_{ext}$ is decreased (increased), the solution will latch onto its nearest neighbor, thereby forcing the Overhauser field to maintain a positive (negative) sign for the duration of the downsweep (upsweep) scan. 

The mode-locked precession frequencies for all of the results are observed to make a switch in precession frequency at the external fields that correspond to the edges of the Overhauser fields' sawtooth profile (see Fig. \ref{fig:2}). However, locations of $B_{ext}$ where the edges of the sawtooth profile occur for depolarizing $B_{ext}$ in comparison to upsweeping and downsweeping $B_{ext}$ are different.

Switching of mode-locked precession frequencies for the depolarized results can be seen to occur around 98 mT, 123 mT, and 147 mT, which are all integer multiples of $B_0$ and located where the Overhauser field goes from being negative to positive in sign. For downsweeping (upsweeping) $B_{ext}$, the switching of precession frequency occur around 111 mT, 136 mT, and 160 mT, which are all half integer multiples of $B_0$ and located where the Overhauser field goes from its maximum (minimum) to zero. Although the locations of $B_{ext}$ where the sawtooth edges occur are the same for downsweeping and upsweeping, an external field-dependent hysteresis behavior is observed in the Overhauser field.

\begin{figure}
    \hspace{-0.5em}\includegraphics[width=1\columnwidth]{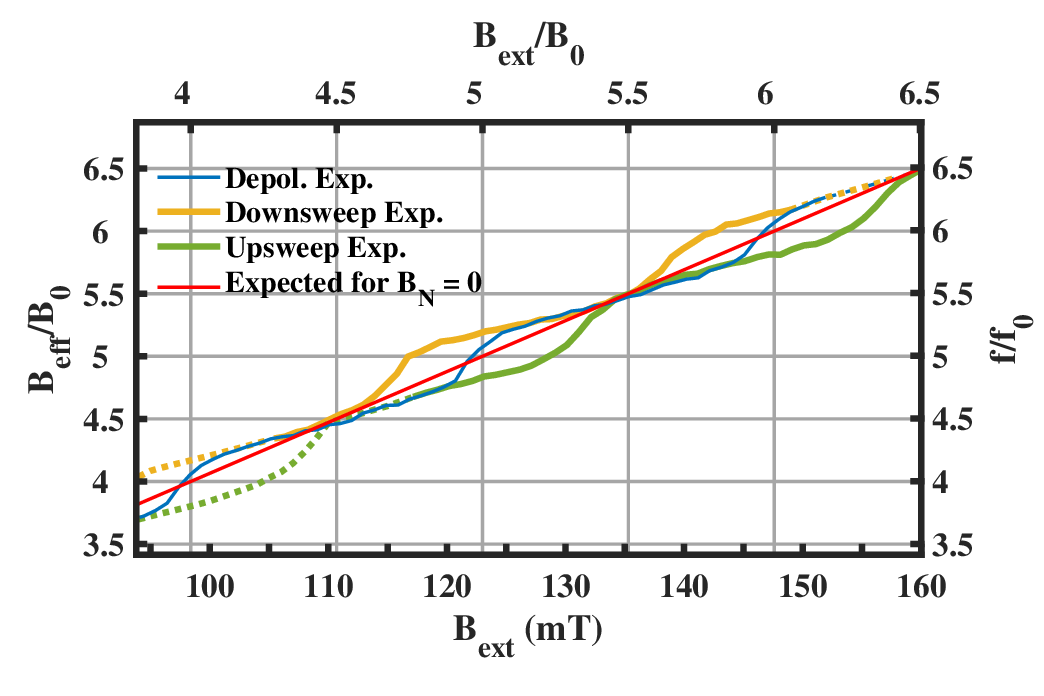}
    \caption{Effective magnetic field $B_{eff}$ (precession frequency $f$) on the left (right) in units of $B_0$ ($f_0$) plotted as a function of $B_{ext}$ ($B_{ext}$ in units of $B_0$) on the bottom (top). The dashed lines are extrapolations of $B_{eff}$ and $f$ using the extrapolated latched solutions from Fig. \ref{fig:3}. The red line depicts the expected results for no presence of an Overhauser field.}
    \label{fig:4}
\end{figure}

A manifestation of this Overhauser field hysteresis is the discretization of bistable mode-locked electron spin precession modes occurring in the same band of $B_{ext}$, which occurs because the nature of the Overhauser field hysteresis leads to two different and discrete effective magnetic fields occurring in the same band of $B_{ext}$. The discreteness of the effective magnetic field is due to the Overhauser field counteracting the change in $B_{ext}$. For example, Fig. \ref{fig:3c} (\ref{fig:3d}) show that the Overhauser field increases (decreases) for a decrease (increase) in $B_{ext}$, thereby counteracting the change in $B_{ext}$ for both downsweeping (upsweeping) and discretizing the effective magnetic field that the electrons experience.

Figure \ref{fig:4} presents the effective field (left vertical axis) as well as precession frequency (right vertical axis) attributed to depolarizing, downsweeping, and upsweeping $B_{ext}$ as well as what would be expected for no presence of an Overhauser field. The bistable mode-locked electron spin precession frequencies are specific to the downsweep and upsweep results, which can be seen to reside above and below the expected precession frequency for no Overhauser field, respectively. This identifies the bistable precession frequencies: the electrons will latch onto a higher and lower order spin precession mode in the same band of $B_{ext}$ for downsweeping and upsweeping, respectively. For example, in the $4.5B_0$ to $5.5 B_0$ band of external field, the electrons will exhibit a predominant higher (lower) precession frequency mode residing between $5f_0$ and $5.5f_0$ ($4.5f_0$ and $5f_0$) for downsweeping (upsweeping) $B_{ext}$. 

Extrapolations of the effective magnetic field and spin precession modes are shown as dashed lines in Fig. \ref{fig:4}. The extrapolations were adopted from the extrapolated latched solutions associated with the downsweep (upsweep) Overhauser fields shown as the color filled circles in Fig. \ref{fig:3c} (\ref{fig:3d}). Considering the periodic nature of the possible solutions to the Overhauser field, the bistable spin precession modes will continue to be bound between external fields of 

\begin{equation}\label{eq:9}
    (k-1/2)B_0\boldsymbol{<B_k<}(k+1/2)B_0
\end{equation}

\noindent with the center of the band being determined by $kB_0$, where $k$ denotes the corresponding mode of interest. Pushing this extrapolation into the precession frequency modes, the higher and lower $k^{th}$ modes -- which are excited by downsweeping and upsweeping $B_{ext}$ -- will have their predominant precession frequency being bound within 

\begin{subequations}\label{eq:10}
\begin{equation}\label{eq:10a}
    kf_0\boldsymbol{<f_k^{+}<}(k+1/2)f_0
\end{equation}
\begin{equation}\label{eq:10b}
    (k-1/2)f_0\boldsymbol{<f_k^{-}<}kf_0
\end{equation}
\end{subequations}

\noindent where $f_k^{+}$ and $f_k^{-}$ correspond to the higher and lower order electron spin precession modes occurring in the same band of $B_k$, respectively. The supplementary materials demonstrates the robustness of this equality by presenting simulations of the spin dynamics for a broader range of the externally applied magnetic field.

In summary, this report experimentally and computationally addresses the discretization of a higher and lower electron spin precession modes defined by Eqs. \ref{eq:10a} and \ref{eq:10b} existing in the same band $B_{ext}$ defined by Eq. \ref{eq:9}. The discretization and mode-locking of the electron spin precession frequency is a result of the electrons experiencing an effective magnetic field that is a sum of $B_{ext}$ and $B_N$ (Eq. \ref{eq:a1}). It is the action of $B_N$ that causes the discretization of the effective magnetic field, thereby focusing the electrons in an ensemble of spins into coherent mode-locked spin precession frequencies. Attaining the higher and lower modes of mode-locked electron spin precession in the same band of $B_{ext}$ is attributed to the hysteretic nature of the Overhauser field. 

Considering that a hysteresis is rooted in the fact that nuclear magnetic moments may occupy the up or down spin state, here, negatively detuned periodic optical pumping and downsweeping (upsweeping) $B_{ext}$ is connected to coherently rotating electron spins parallel (antiparallel) to the direction of $B_{ext}$. The parallel (antiparallel) rotation of electron spins therefore result in electrons coupling to the nuclei through the hyperfine interaction, thereby preparing spin up (down) polarized nuclei, which has a correspondence to the nuclear magnetic moments. The spin up (down) polarized nuclei will dominate nuclear spin occupancy over the opposing type, which manifests as the Overhauser field maintaining a positive (negative) sign for downsweeping (upsweeping) $B_{ext}$ as demonstrated in Fig. \ref{fig:3c} (\ref{fig:3d}). 

Although the magnitude of the Overhauser fields presented in Fig. \ref{fig:3} correspond to a modest polarization of the nuclear ensemble -- which have been reported to be on the order of $\sim $ 10\% for systems similar this report \cite{NuclearPol1, NuclearPol2} -- that has too low of nuclear state purity for solely utilizing the nuclei toward quantum information processing, the controlled tuning of the electron spin precession frequency due to the modest Overhauser fields are useful for conceptualizing other spin-based information processing protocols. Furthermore, considering the theoretical demonstrations of nuclear spin squeezing and entanglement in very similar systems \cite{Squeeze1, Squeeze2}, future works may use these reported bistable Overhauser fields -- which manifests as different signs of $B_N$ attained by upsweeping and downsweep $B_{ext}$ -- existing in the same band of $B_{ext}$ to investigate how bistable nuclear spin states effect nuclear spin squeezing and entanglement.

\begin{acknowledgments}
E.K. was supported by the University of Michigan’s Rackham Science Fellowship. This material is based upon the work supported by the National Science Foundation under Grant No. DMR-2207162.
\end{acknowledgments}

\end{document}